\documentclass[graphicx,reprint]{revtex4-1}

\usepackage{xcolor}
\usepackage{sistyle}
\usepackage{gensymb}
\usepackage{graphicx}

\begin{document}

\title{High Resolution Viscosity Measurement by Thermal Noise Detection}

\author{Felipe Aguilar Sandoval}
\email{felipe.aguilarsan@usach.cl}
\affiliation{Departamento de F\'isica, Universidad de Santiago de Chile, Avenida Ecuador 3493, 9170124 Estaci\'on Central, Santiago, Chile}
\author{Manuel Sep\'ulveda}
\affiliation{Departamento de F\'isica, Universidad de Santiago de Chile, Avenida Ecuador 3493, 9170124 Estaci\'on Central, Santiago, Chile}
\author{Ludovic Bellon}
\affiliation{ Laboratoire de Physique, Universit\'e de Lyon, Ecole Normale Sup\'erieure de Lyon, CNRS, 46 All\'ee d'Italie, F-69007 Lyon, France}
\author{Francisco Melo}
\email{francisco.melo@usach.cl}
\affiliation{Departamento de F\'isica, Universidad de Santiago de Chile, Avenida Ecuador 3493, 9170124 Estaci\'on Central, Santiago, Chile}

\begin{abstract}
An interferometric method is implemented in order to accurately assess the thermal fluctuations of a micro-cantilever sensor in liquid environments.  The power spectrum density (PSD) of thermal fluctuations together with Sader's model of the cantilever allow for the indirect measurement of the liquid viscosity with good accuracy.  The good quality of the deflection signal and the characteristic low noise of the instrument allow for the detection and corrections of drawbacks due to both, the cantilever shape irregularities and the uncertainties on the position of the laser spot at the fluctuating end of the cantilever.   
Variation of viscosity below the \SI{0.03}{mPa.s} were detected with the alternative to achieve measurements with a volume as low as \SI{50}{\mu L}.
\end{abstract}

\maketitle

\section{Introduction}

The measurement of rheological properties like fluid viscosity is of common interest in several science fields, ranging from engineering to biology, with applications including process control~\cite{Yablon-2014, Kress-Rogers-2001, Harrison-2006, Cao-Paz-2012} and fluids analysis for diseases diagnosis~\cite{Boss-2011, Cakmak-2013, Barnett-1958}.   On the other hand, the atomic force microscopy (AFM) has became a valuable technique that is applied transversely to scientific disciplines.  AFM can be used in diverse modes to image either hard or soft surfaces in both dry and fluid environment    
~\cite{Jalili-2004, Dufrene-2013}.  Besides atomic resolution imaging,  force spectroscopy has emerged as a powerful tool to characterize the mechanics of soft materials at the nano scale ~\cite{Butt-2005, Hodges-2002, Neuman-2008} and single molecule level ~\cite{Kadaa-2008}.

Moreover, a new class of miniaturized, highly sensitive biosensors, has been envisioned based on the high sensitivity of micro fabricated silicon cantilevers to bending ~\cite{Lavrik-2004, Tamayo-2012}.   The selective adsorption of  molecules onto the cantilever surface induces surface stresses, which ultimately leads to a deflection signal.    Thermal noise detection is a less known and promising technique that has proven useful for the characterization of the cantilever mechanics \cite{Bechhoeffer-1993} and its surrounding media ~\cite{Butt-1995, Stark-2001}.  

In order to explore advantages of cantilevers as sensors for fluid properties in applications in which only small amount of material is available and where a good time response is required, we developed a device able to measure thermal fluctuations of a cantilever in the testing fluid environment, with high accuracy.   We extended the capabilities of a previously developed interferometric AFM working in gases ~\cite{Paolino-2013}  to liquids environment.  The analysis of the power spectrum density (PSD) of the fluctuating cantilever-deflection, through the Sader's model~\cite{Sader-1998}, allows for the measurement of the fluid viscosity.

Previous works have used Sader's model and the optical lever method ~\cite{Butt-2005} for detecting deflections and deducing fluid viscosity ~\cite{Bergaud-2000, Papi-2006, McLoughlin-2007, Hennemeyer-2008, Paxman-2012}.  In general, the information obtained was through a partial fit of the PSD spectra (p.e. of the first resonance) ~\cite{McLoughlin-2007} or by solving an approximated equation system to link the frequency shift and the quality factor to the fluid viscosity and density ~\cite{Hennemeyer-2008}, which ultimately leaded to accuracy of about $5\%$.  In this article, we present the advantage of an interferometric method~\cite{Paolino-2013} to achieve improved PSD measurements, in order to obtain a reliable fit of the whole spectrum with Sader's model for the two first normal modes of the cantilever immersed in a liquid, which adds a quality test to the fit.  This procedure simultaneously allows us for the absolute measurement of the viscosity.  In addition, the micro-cantilever is actuated only by thermal noise,  reducing the equipment needed to implement our method, in comparison with externally-excited cantilever methodology.\footnote{This work uses only thermal noise instead of an external cantilever excitation.   However, highly performance external devices~\cite{Bonfig-1988,Matko-2009,Matko-2011} for cantilever excitation combined with our interferometer is an interesting avenue to explore in order to reduce or compensate temperature influence and other undesirable effects.} Our sensing configuration requires small amount of fluid, ranging from \SI{1}{mL}, for full immersion of the cantilever holder, to a drop of \SI{50}{\mu L} that is the minimum amount filling the gap between the cantilever and the optical window.  This last option is well suited for expensive and scarce fluids and biological samples ~\cite{Hennemeyer-2008}.

The article is structured as follows, in section \ref{section:Theory} we describe the model used to fit our experimental data. Section \ref{section:Setup} presents improvement of the interferometer described in~\cite{Paolino-2013} to extend its application in liquids media. In section \ref{section:Methods} we describe the methodology and the protocols to prepare testing fluids and to perform measurements.  Section \ref{section:Results}  presents the obtained values of viscosity and comparisons with reference values. Finally, section \ref{section:Discussions}  discusses the results validity, compares the model with the experimental PSD and considers the effect of the microcantilever geometry, previous to the general conclusions of this work.

\begin{figure}%
\centering
\includegraphics{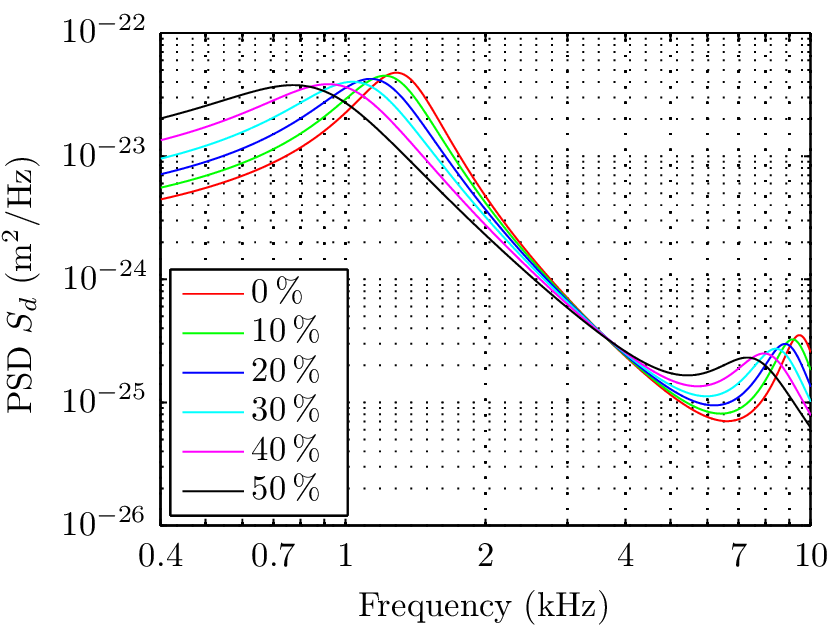}%
\caption{Theoretical curves of PSD of thermal noise induced deflection of a cantilever for Sader's model for different glycerol concentrations, at \SI{24}{\degree C}; the corresponding viscosities and densities are listed in Table~\ref{teo_val}. Effects on the PSD amplitude, width, and resonant frequencies are clearly visible. The geometry of the cantilever for this modeling is a length $L=\SI{350}{\mu m}$, a width $W=\SI{32.5}{\mu m}$ and a thickness $H=\SI{1.0}{\mu m}$.}%
\label{sim_modes}%
\end{figure}

\section{Theory \label{section:Theory}}

According to Sader's theory\cite{Sader-1998} and Bellon's article\cite{Bellon-2008}, the PSD for each normal mode of the deflection $d$ for a microcantilever immersed into a thermal bath at temperature $T$ is

\begin{equation}
S_{d_{n}}(\omega)=\frac{2k_{B}T}{\pi}\frac{\gamma_{\mathrm{eff}}(\omega)}{(k_{n}-m_{\mathrm{eff}}(\omega)\omega)^2+(\gamma_{\mathrm{eff}}(\omega)\omega^{2})^2}
\label{eq:sader_nm}
\end{equation}

where $d_{n}$ is the amplitude of the $n-$th normal mode of the cantilever and a fluctuating variable due to the thermal bath, $\gamma_{\mathrm{eff}}(\omega)$ is the frequency dependent effective  damping coefficient, $m_{\mathrm{eff}}(\omega)$ is the frequency dependent effective mass, $k_{n}$ is the stiffness of the $n-$th mode, $k_{B}$ is the Boltzmann constant and $\omega$ is the angular frequency. The effective values $m_{\mathrm{eff}}$ and $\gamma_{\mathrm{eff}}$ are

\begin{eqnarray}
m_{\mathrm{eff}}(\omega)&=&m+m_{f}\Gamma_{r}(\omega) \label{eq:meff} \\
\gamma_{\mathrm{eff}}(\omega)&=&m_{f}\omega\Gamma_{i}(\omega)
\label{eq:gammaeff}
\end{eqnarray}

where $m$ is the mass of the cantilever, $m_{\mathrm{f}}=\pi\rho LW^{2}/4$ is the mass of the cylinder of fluid of density $\rho$ surrounding the cantilever of length $L$ and width $W$ and $\Gamma=\Gamma_{r}+i\Gamma_{i}$ is the hydrodynamic function in the Sader's model context, whose values are dependent of the fluid density and viscosity.
For the sake of completeness, we introduce explicitly the hydrodynamic function for a rectangular beam

\begin{equation}
\Gamma(\omega)=\Omega(\omega)\Gamma_{\mathrm{circ}}(\omega)
\label{eq:Gamma}
\end{equation}

where $\Gamma_{\mathrm{circ}}$ is the hydrodynamic function for a cylinder expressed as

\begin{equation}
\Gamma_{\mathrm{circ}}(\omega)=1+\frac{4iK_{1}(-i\sqrt{i\mathrm{Re}})}{\sqrt{i\mathrm{Re}}K_{0}(-i\sqrt{i\mathrm{Re}})}
\label{eq:Gammacirc}
\end{equation}

where $K_{0}$ and $K_{1}$ are modified Bessel functions of the third kind and Re is the Reynolds number, expressed as

\begin{equation}
\mathrm{Re}=\frac{\rho W^{2}\omega}{4\eta}
\label{eq:Re}
\end{equation}

where $\rho$ is the density of the fluid where the cantilever is immersed, $W$ is the width of the cantilever and $\eta$ is the viscosity of the fluid. $\Omega(\omega)$ is a correction function, expressed in the real and imaginary parts as

\begin{eqnarray}
\Omega_{r}(\omega)&=&(0.91324-0.48274r+0.46842r^{2}-0.12886r^{3} \nonumber \\
&+&0.044055r^{4}-0.0035117r^{5}+0.00069085r^{6}) \nonumber \\
&\times&(1-0.56964r+0.48690r^{2}-0.13444r^{3} \nonumber \\
&+&0.045155r^{4}-0.0035862r^{5} \nonumber \\
&+&0.00069085r^{6})^{-1} \label{eq:OmegaR} \\
\Omega_{i}(\omega)&=&(-0.024134-0.029256r+0.016294r^{2} \nonumber \\
&-&0.00010961r^{3}+0.000064577r^{4} \nonumber \\
&-&0.000044510r^{5})\times(1-0.59702r+0.55182r^{2} \nonumber \\
&-&0.18357r^{3}+0.079156r^{4}-0.014369r^{5} \nonumber \\
&+&0.0028361r^{6})^{-1} \label{eq:OmegaI} \\
r&=&\log_{10}(\mathrm{Re}) \label{eq:r}
\end{eqnarray}

where $\Omega(\omega)=\Omega_{r}+i\Omega_{i}$. The physical origin of the hydrodynamic function can be found in Sader's work \cite{Sader-1998}. It is worth nothing that in Sader's model the dissipation is assumed homogeneous along the cantilever. If in addition, the thermal noise is uncorrelated on different normal modes \cite{Bellon-2008}, the PSD of the deflection $d$ is written as

\begin{equation}
S_d(x,\omega)=\sum^{\infty}_{n=1}S_{d_{n}}(\omega)\left|\phi_{n}(x)\right|^{2}
\label{eq:PSD_d}
\end{equation}

where $\phi_{n}$ are the basis of normal modes and $x$ is the spatial coordinate along the cantilever length (the position where the deflection is measured).

As an illustrative example we plot the PSD of the deflection of the cantilever  (Fig.~\ref{sim_modes}) using  Eq. ~\ref{eq:sader_nm} and the geometric parameters of the cantilever described in the experimental section. The PSD is computed for different values of viscosities and densities obtained for water solutions of different concentrations of glycerol that are listed in Table~\ref{teo_val} at \SI{24}{\degree C} and computed through the work of Cheng et al~\cite{Cheng-2008}.  A frequency shift of the resonance peaks together with significant broadening and amplitude variation are clearly observed, illustrating the sensitivity of the PSD to viscosity variations.

\begin{table}%
\centering
\begin{tabular}{ccr}
$[G]_{v}$ \SI{}{(\%v/v)}&$\eta_{\mathrm{ref}}$ \SI{}{(mPa.s)}&$\rho_{\mathrm{ref}}$ \SI{}{(kg/m^{3})}\\
\hline
\hline
0&0.9135&997.1\\
10&1.2501&1029.7\\
20&1.7784&1060.6\\
30&2.6529&1090.0\\
40&4.1971&1118.0\\
50&7.1505&1144.7
\end{tabular}
\caption{Reference values~\cite{Cheng-2008} of viscosity $\eta_{\mathrm{ref}}$ and density $\rho_{\mathrm{ref}}$ are listed for different volume concentrations of glycerol $[G]_{v}$ used at a temperature of \SI{24}{\degree C}.}
\label{teo_val}
\end{table}

\section{Setup \label{section:Setup}}

\subsection{Sensing}

\begin{figure}%
\centering
\includegraphics[width=0.5\columnwidth]{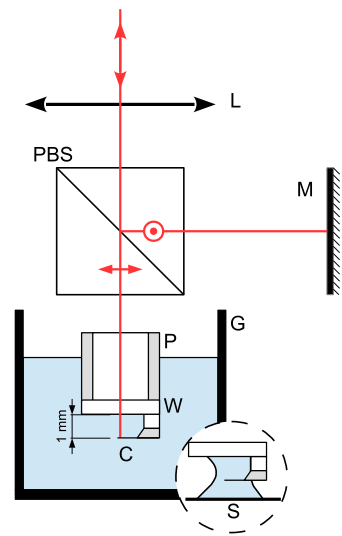}
\caption{Using a polarized beam splitter (PBS), the input beam is divided in two, with crossed polarizations as shown in the divided laser beam. An external mirror (M) reflects the reference beam and provides fine control of the overlap of the returned beams. A lens (L) focuses the probe beam at the free end of the cantilever (C). The reflected beams, probe and reference, are superposed after the PBS. Interferences are obtained after projecting the initial polarizations in the analysis area (not shown here, see ref.~\cite{Paolino-2013} for details. The sensing beam path is protected by a stainless steel pipe (P).  An optical quality window (W)  ensures the transition from air to liquid with minimum refraction and undesirable reflections in order to reach the free end of the cantilever (C). The liquid is contained in a Pyrex glass (G).   Inset: the micro droplet configuration  uses a microscope slide  to support a liquid drop (\SI{50}{\mu L}).  The micro-cantilever is fully immersed and the space between the window (W) and the microscope slide (S) is filled with the fluid, forming a meniscus. In order to prevent beam distortion, the window W is of laser quality with suitable antireflection coating.}%
\label{q_conf}%
\end{figure}

Our device is an interferometric AFM, originally designed for nano-mechanical measurements of thermal noise in gaseous environment.  Working principle and a detailed description of the interferometer can be found in ref. \cite{Paolino-2013}. 
The principle applied for sensing deflection is of the interferometric kind; a stable reference beam is located outside of the fluid cell using a Michelson like polarized set-up.  An external mirror (M) reflects the reference beam and provides fine control of the overlap of the returned beams. A lens (L) whose fine position is controlled through a motorized three-axis system focuses the probe beam (z-control) and allows to position the resulting laser spot at the free end of the cantilever (C) with good accuracy.  This configuration allows for the fine tuning maximizing the contrast signal significantly (Fig. \ref{q_conf}). This also help to minimize the undesirable effects of the refraction at the interfaces due to small misalignment of the probe beam.  The higher resolution reached with the interferometric detection is due to null dependence of the sensitivity with respect to the phase difference between the reference and probe beam.  Thus, the background noise is limited just by the shot noise of the photodiodes.   Other advantage is the deflection detection range that can reach several microns with suitable functions in the data acquisition system.  This feature is due to an automatized system detecting the unwrapped phase.

\subsection{Immersion area and fluid cell}

The cantilever holder for immersion in liquids is a small stainless steel pipe closed at one end by a laser quality coated window that is glued with optical UV adhesive (NOA 81, Norland Products Inc), (Fig. \ref{q_conf}). This configuration provides a suitable transition from air to liquid, avoiding drawbacks due to refraction and undesirable reflections.  The whole interferometer (including the components of the Fig.~\ref{q_conf}) is mounted on a manual translation microscope turret, which provides vertical motion and allows introducing of the micro-cantilever into the liquid gently.  The interferometer is then moved downward until the fluid has filled the gap between the cantilever and the optical window.
An alternative configuration is obtained by  replacing the glass cell G (Fig. \ref{q_conf}) with a microscope slide to locate a drop ($\sim$\SI{50}{\mu L}) of fluid sample just below the cantilever  (see inset of Fig. \ref{q_conf}).  This option is recommended for scarce liquids and biological fluids available only in small quantities.

\section{Methods \label{section:Methods}}

The cantilever used for our experiments has a rectangular geometry, with nominal values of $L=\SI{350}{\mu m}$ in length, $W=\SI{33.5}{\mu m}$ in width and $H=\SI{1.0}{\mu m}$ in thickness (MikroMasch, HQ:CSC38/Cr-Au).  Notice that the nominal width is the average across the thickness because the cantilever cross section is trapezoidal. 
For a fine tuning of the mechanical properties and geometry of the cantilever, we consider as an input and absolute values  of viscosity and density those of mQ water.  Indeed, with this fixed value of viscosity and fluid density, through  small adjustments of the cantilever geometry, length, width, thickness in the model we reproduce the PSD obtained experimentally with pure water.  This procedure allows to find effective values featuring the cantilever geometry by correcting uncertainties due to irregularities of the geometrical form of the cantilever.  In addition, it is important to mention that the position of the cantilever with respect to the spot of the sensing laser can be determined by scanning the cantilever with the same laser spot, using a micro positioning system.  Thus, the laser spot can be positioned at the desired position $x$ near the free end of the cantilever ($x=L$).   Notice that knowing the exact point on the cantilever at which thermal fluctuations are detected is crucial for a good repeatability and accuracy in the fit procedure because the theoretical PSD is calculated for this particular point at the cantilever.  We realized that fine tuning of this parameter improved the fit quality, therefore it was allowed to slightly vary in the fitting procedure.

As reference fluids we use different volume concentrations of glycerol, from \SI{1}{\%} to \SI{50}{\%}. The density $\rho_{\mathrm{ref}}$ and the viscosity $\eta_{\mathrm{ref}}$ values used as reference data were computed using an empirical equation \cite{Cheng-2008}.  The volume of each reference fluid used in the experimental cell was \SI{1}{mL}.  When performing the measurements the temperature variation from run to run was corrected.  To prevent rapid temperature variations, the whole setup was enclosed in an acrylic box and the fluid was allowed to thermalize prior measurements. Two thermometers in the acrylic box allowed for the temperature monitoring.  A digital thermometer with a resolution of \SI{0.1}{\degree C} registered temperature in the enclosure whereas a thermocouple located in contact with the glass window (W) monitored the sample temperature. We whole variation of temperature during the experiments was kept smaller than \SI{0.1}{\degree C}.

Two sets of reference fluids were prepared with two different protocols. The first group of fluids ranges from \SI{1}{\%} to \SI{10}{\%} of glycerol and is obtained from a seed dilution of \SI{10}{\%} glycerol by gradual dilution in steps of \SI{1}{\%}.  The second group includes concentrations from \SI{10}{\%} to \SI{50}{\%} in steps of  \SI{10}{\%} and were prepared one by one. All mixtures were prepared with mQ water.

\section{Results \label{section:Results}}

\begin{figure}%
\centering
\includegraphics{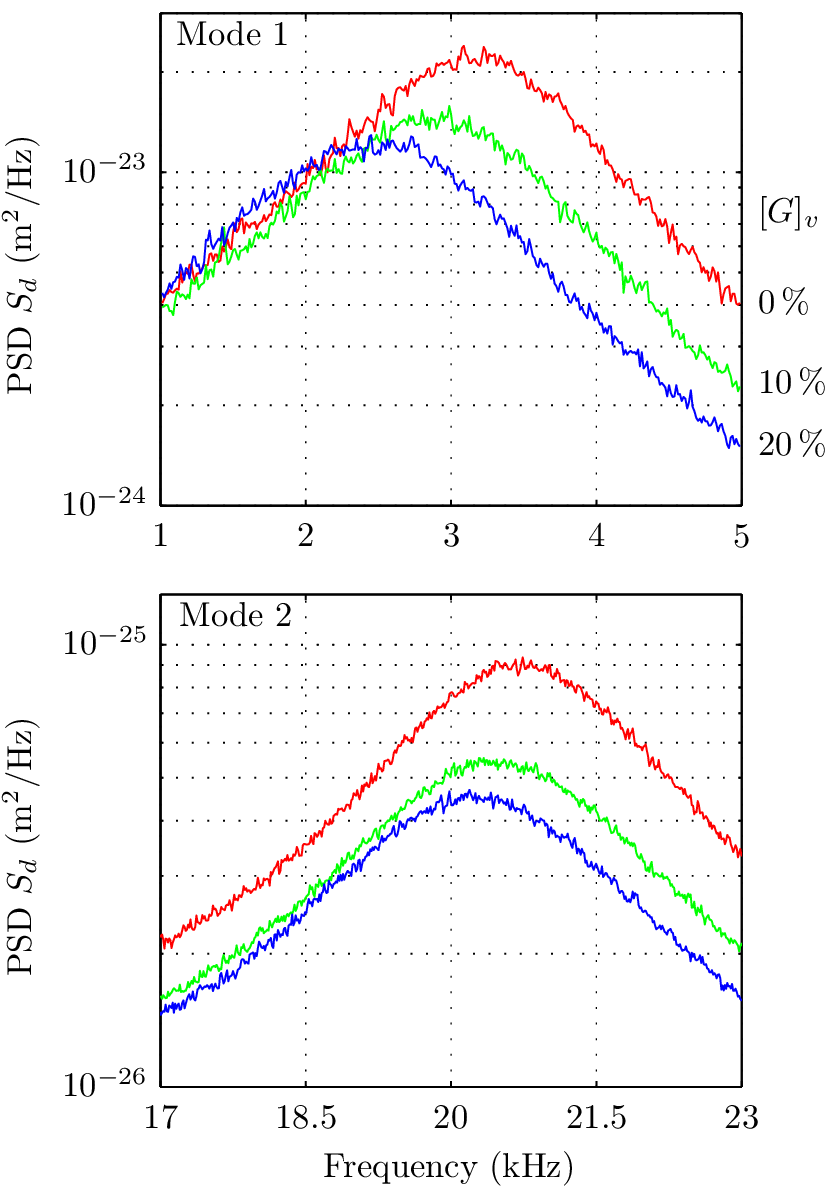}%
\caption{Zoom around the 2 first resonances of the thermal noise spectra of the deflection of a cantilever immersed in water-glycerol mixtures. These PSDs show increasing frequency shifts and resonances widening for increasing glycerol concentration.}%
\label{lh_v}%
\end{figure}

\begin{figure}%
\centering
\includegraphics{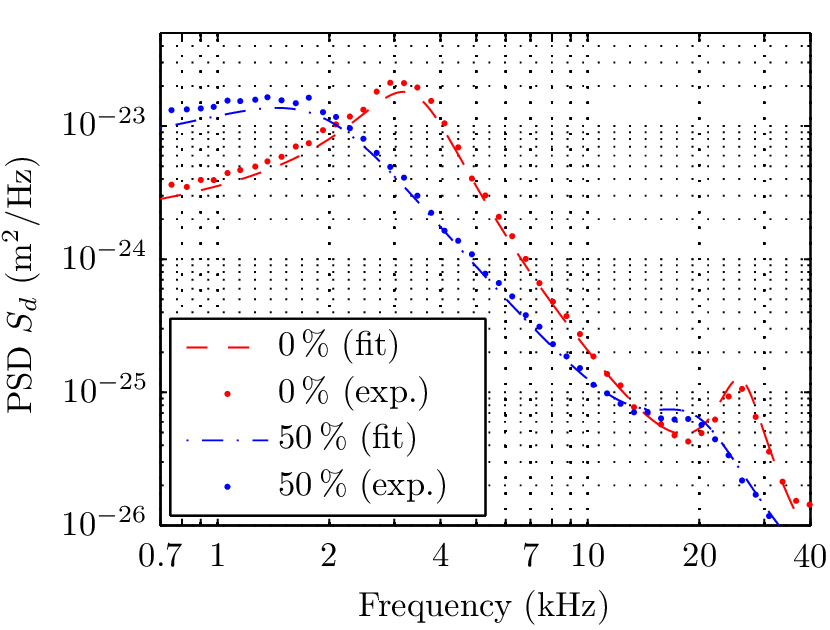}%
\caption{Experimental data (dots) and fits with Sader's model (dashed lines)  for the PSD of thermal noise induced deflection for pure water (mQ) and a solution of $50\%$ of glycerol.}%
\label{f_ex}%
\end{figure}

The measurements were divided by concentration in two groups of testing fluids;  one from $0\%$ to $9\%$ glycerol in steps of $1\%$ and the other from $10\%$ to $50\%$ glycerol in steps of $10\%$.  The experimental PSD of the deflection for different concentrations (Fig.~\ref{lh_v}) show that the effect of viscosity variations produces both a noticeable frequency shift and a resonance widening for increasing glycerol concentration.  The quality of the fitting for the whole PSD is shown in 
 Fig.~\ref{f_ex}. 

For the first group (concentrations from pure water to \SI{9}{\%} of glycerol), the viscosities $\eta_{\mathrm{exp}}$ obtained by fitting the PSD are compared to the reference values $\eta_{\mathrm{ref}}$ in Table~\ref{tab_low_v}.   Errors up to \SI{1}{\%} are obtained, although in most cases errors are much smaller, with a mean value of \SI{0.5}{\%} for this first group of measurements.   We conclude that our procedure allows for the detection of viscosity variations as low as \SI{1}{\%}, differentiating changes bellow \SI{0.03}{mPa.s}.   Values of cantilever's geometry obtained from the fitting procedure, width $W_f$ and length $L_f$,  are $L_f=354\mu$m, $W_f=32.9\mu$m, which compare well to the nominal values, $L=359\mu$m and $W=33.5\mu$m.  

\begin{table}%
\centering
\begin{tabular}{cccr}
$[G]_{v}$ \SI{}{(\%v/v)}&$\eta_{\mathrm{ref}}$ \SI{}{(mPa.s)}&$\eta_{\mathrm{exp}}$ \SI{}{(mPa.s)}&error \SI{}{(\%)}\\
\hline
\hline
0&0.913&0.912&-0.07\\
1&0.938&0.940&0.16\\
2&0.967&0.957&-1.11\\
3&0.995&0.993&-0.23\\
4&1.026&1.029&0.30\\
5&1.058&1.062&0.38\\
6&1.092&1.096&0.35\\
7&1.125&1.125&0.04\\
8&1.162&1.177&1.34\\
9&1.200&1.212&0.98\\
\end{tabular}
\caption{Comparison between reference viscosities $\eta_{\mathrm{ref}}$ and those extracted from the fits of the PSDs of the measured thermal noise $\eta_{\mathrm{exp}}$, for mixtures of water and glycerol in the range of 0 to \SI{10}{\%}.}
\label{tab_low_v}
\end{table}

In turn, for concentrations from \SI{10}{\%} to \SI{50}{\%} of glycerol, the measured viscosity has bigger errors with  respect to the reference value, as seen in Table~\ref{tab_high_v}.  Such increasing errors are expected and demonstrate that our procedure for tuning cantilever parameters (performed with pure water)  is valid for about one order of magnitude variation in viscosity with a maximum error less than \SI{10}{\%}.   
 
\begin{table}%
\centering
\begin{tabular}{cccr}
$[G]_{v}$ \SI{}{(\%v/v)}&$\eta_{\mathrm{ref}}$ \SI{}{(mPa.s)}&$\eta_{\mathrm{exp}}$ \SI{}{(mPa.s)}&error \SI{}{(\%)}\\
\hline
\hline
10&1.256&1.278&1.7\\
20&1.783&1.785&0.1\\
30&2.661&2.580&-3.0\\
40&4.197&4.047&-3.6\\
50&7.150&6.573&-8.1\\
\end{tabular}
\caption{Comparison between reference viscosities $\eta_{\mathrm{ref}}$ and those extracted from the fits of the PSDs of the measured thermal noise $\eta_{\mathrm{exp}}$, for mixtures of water and glycerol in the range of 10 to \SI{50}{\%}.}
\label{tab_high_v}
\end{table}

\begin{figure}%
\centering
\includegraphics{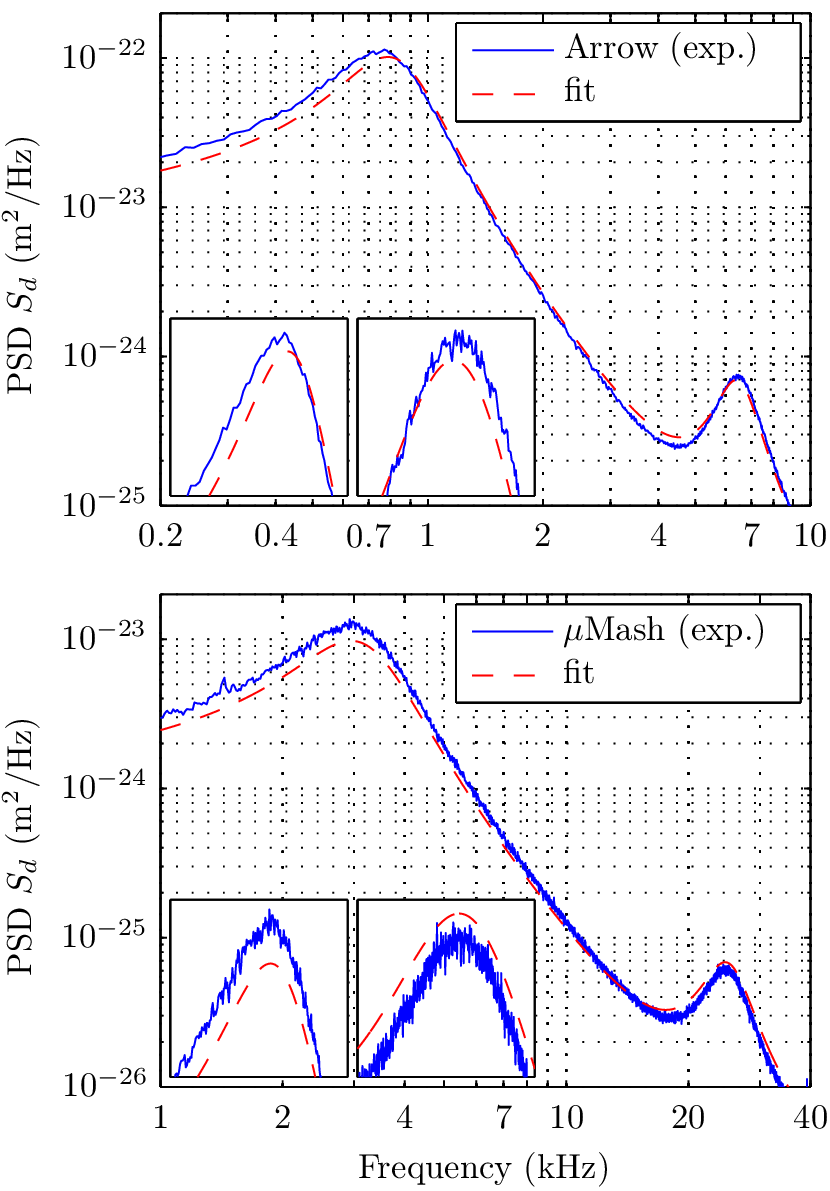}%
\caption{Comparisons of the quality of the fits of the PSDs for both cantilevers used. (top) The Arrow TL1, NanoWorld cantilever produces an optimized fit that has a notorious mismatch in the resonances frequencies.  Insets: details of the fits for the first and the second cantilever resonance. (bottom) Same as top curve for the MikroMasch (HQ:CSC38) cantilever. Insets:  the fit with the Sader's model captures both resonances simultaneously.}%
\label{comp_sader}%
\end{figure}

\section{Discussions \label{section:Discussions}}

\begin{figure}%
\centering
\includegraphics[width=0.5\columnwidth]{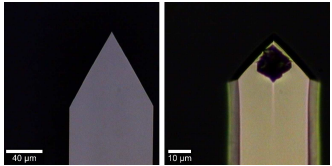}%
\caption{ Calibrated images of the end of the cantilevers. (left) Arrow TL1 cantilever, (right)  MikroMasch cantilever.}%
\label{levers}%
\end{figure}

In  additional experiments we tested the performance of a different cantilever, the Arrow TL1 (NanoWorld, without coating), of nominal values $L=$ \SI{500}{\mu m},  $W=$\SI{100}{\mu m} and \SI{1}{\mu m} thickness. The viscosity values obtained with this cantilever (not presented here), under the same conditions, differ from those obtained with MikroMasch.  First, we observed that when the nominal values of W and L are used, viscosities values resulted unrealistic, revealing an inaccuracy.  Second, when W and L are used as free parameters to adjust the PSD of pure water with the value of pure water viscosity as an input, the equivalent L and W (\SI{471}{\mu m} and \SI{77}{\mu m} respectively) differed from nominal values,  confirming the previous inaccuracy.  With these parameters, the measured values of viscosities differed about 4\% respect to the reference ones (for the set 1 to 10\% glycerol).   In addition, we detected a mismatch between two adjacent modes with respect to the model used (Fig.~\ref{comp_sader}a).  It was not possible to fit both resonances with a single set of parameters, which further questions the trust of the fit.  Our experiments suggest that these effects are due to the size of triangular head (about \SI{85}{\mu m}) of  the Arrow TL1 relative to its total length.   Indeed,  the MikroMasch cantilever also has a triangular end, however its size is only about \SI{20}{\mu m} which is small compared to the cantilever total length. Fig.~\ref{levers} show  pictures for comparison of the free ends of both cantilevers.  We concluded that the effect of the shape of the cantilever end is reduced by considering large aspect ratio cantilevers.  In the case of MikroMasch cantilever,  a fine tuning of  W and L allows for reducing errors in viscosity from 3\% to 1\%. 
 
In spite of these subtle geometrical effects, that could offer opportunities for further improvements in the use of micro cantilever for viscosity measurements,  our methodology allows for detection of viscosity changes as low as 1\%.  The errors relative to the reference values increase when the changes of viscosity are greater than \SI{30}{\%}. 


%
%
%
%
%


%
%
%
%
%
%
%


\section{Conclusions \label{section:Conclusions}} 

The interferometric detection for cantilever deflection enables the measurement of the PSD of thermal fluctuations with a very high resolution and low noise. This is useful to discriminate small changes of the viscosity of the surrounding media of the cantilever in less than \SI{0.03}{mPa.s}. The amount of liquid needed to perform the measurements is an additional advantage of the method, allowing volumes as small as \SI{50}{\mu L}.  Regarding the time resolution of the method,  measurements as fast as four data point per minute have been achieved,  which suggests that the methodology is well adapted to follow time evolution of fluid viscosity, due to either structural changes in complex fluids or variations of some external parameter.  However, the fluid under scope must be sufficient transparent and free of impurities producing scattering  and perturbation of the probe beam.

The geometry of the micro-cantilever used was shown to play an important role, we observed that the triangular end of the cantilever can modify significantly its equivalent dimensions and  the relation between their resonance frequencies as the model no longer capture accurately the resonances frequencies simultaneously.  Thus, our results suggest that further improvements in fluid properties measurements can be made through the optimization of the cantilevers geometry. 


\acknowledgments

We thank G. Martin for technical assistance and A. Petrosyan for valuable suggestions. This work has been supported by Fondecyt, Chile, under Grant $N^0 1130922$ and Proyecto DICYT $041331$ MHPOSTDOC. 


\section*{Author Contributions}

F. Melo and F. Aguilar and L. Bellon developed the setup to measure cantilever's fluctuations in fluids. M. Sep\'ulveda and F. Aguilar performed experiments and data analysis.  L. Bellon developed optimized scripts for data analysis. All authors contributed equally to the paper writing.

\bibliography{viscosity_revtex}

\end{document}